\documentclass{article}

\usepackage{booktabs}
\usepackage{graphics}
\usepackage{graphicx}
\usepackage{listings}
\usepackage{nameref}
\usepackage{algorithm}
\usepackage{algpseudocode}
\algrenewcommand\algorithmicrequire{\textbf{Input:}}
\algrenewcommand\algorithmicensure{\textbf{Output:}}
\usepackage{bm}
\usepackage{amsmath}
\usepackage{amssymb}
\usepackage{caption} 
\usepackage{multirow}
\usepackage{upgreek}
\usepackage{color}
\usepackage[table]{xcolor}

\usepackage{arxiv}

\usepackage[utf8]{inputenc} 
\usepackage[T1]{fontenc}    
\usepackage{hyperref}       
\usepackage{url}            
\usepackage{booktabs}       
\usepackage{amsfonts}       
\usepackage{nicefrac}       
\usepackage{microtype}      
\usepackage{lipsum}
\usepackage{graphicx}
\graphicspath{ {./images/} }

\begin{document}
\title{GiantHunter: Accurate detection of giant virus in metagenomic data using reinforcement-learning and Monte Carlo tree search}

\author{
 Fuchuan Qu \\
  Dept. of Electrical Engineering\\
  City University of Hong Kong\\
  Kowloon, Hong Kong SAR, China\\
  \And
 Cheng Peng \\
  Dept. of Electrical Engineering\\
  City University of Hong Kong\\
  Kowloon, Hong Kong SAR, China\\
  \And
   Jiaojiao Guan \\
  Dept. of Electrical Engineering\\
  City University of Hong Kong\\
  Kowloon, Hong Kong SAR, China\\
    \And
   Donglin Wang \\
  Sch. of Environmental Science and Engineering\\
  Shandong University\\
  Jinan, Shandong, China\\
  \And
 Yanni Sun \\
  Dept. of Electrical Engineering\\
  City University of Hong Kong\\
  Kowloon, Hong Kong SAR, China\\
  \And
 Jiayu Shang \\
  Dept. of Information Engineering\\
  Chinese University of Hong Kong\\
  New Territory, Hong Kong SAR, China\\
}

\maketitle
\begin{abstract}
\textbf{Motivation:}  Nucleocytoplasmic large DNA viruses (NCLDVs) are notable for their large genomes and extensive gene repertoires, which contribute to their widespread environmental presence and critical roles in processes such as host metabolic reprogramming and nutrient cycling. Metagenomic sequencing has emerged as a powerful tool for uncovering novel NCLDVs in environmental samples. However, identifying NCLDV sequences in metagenomic data  remains challenging due to their high genomic diversity, limited reference genomes, and shared regions with other microbes. Existing alignment-based and machine learning methods struggle with achieving optimal trade-offs between sensitivity and precision. 

\textbf{Results:} In this work, we present GiantHunter, a reinforcement learning-based tool for identifying NCLDVs from metagenomic data. By employing a Monte Carlo tree search strategy, GiantHunter dynamically selects representative non-NCLDV sequences as the negative training data, enabling the model to establish a robust decision boundary. Benchmarking on rigorously designed experiments shows that GiantHunter achieves high precision while maintaining competitive sensitivity, improving the F1-score by 10\% and reducing computational cost by 90\% compared to the second-best method. To demonstrate its real-world utility, we applied GiantHunter to 60 metagenomic datasets collected from six cities along the Yangtze River, located both upstream and downstream of the Three Gorges Dam. The results reveal significant differences in NCLDV diversity correlated with proximity to the dam, likely influenced by reduced flow velocity caused by the dam. These findings highlight GiantHunter's potential to advance our understanding of NCLDVs and their ecological roles in diverse environments.

\textbf{Availability:} The source code of GiantHunter is available via: \href{https://github.com/FuchuanQu/GiantHunter}{https://github.com/FuchuanQu/GiantHunter}.

\textbf{Contact:} \href{yannisun@cityu.edu.hk}{yannisun@cityu.edu.hk}, \href{jiayushang@cuhk.edu.hk}{jiayushang@cuhk.edu.hk}
\end{abstract}

\section{Introduction}
Nucleocytoplasmic large DNA viruses (NCLDVs), also known as "giant viruses", belong to the viral phylum \textit{Nucleocytoviricota}. NCLDVs gained widespread attention with the discovery of Mimivirus, which was initially mistaken for a bacterium due to its large hairy appearance and positive Gram-stain result \cite{scola2003giant, yamada2011giant}. NCLDVs are remarkable for their size: virions can reach up to 1.5 $\upmu$m, and genomes can be as large as 2.5 Mb, rivaling the size and complexity of many bacteria and archaea \cite{aylward2021phylogenomic, koonin2010origin}. Their presence challenges traditional views of viral simplicity. Marker gene surveys reveal that NCLDVs are highly diverse and widely distributed across various environments, including oceans \cite{mihara2018taxon, hingamp2013exploring, monier2008taxonomic}, freshwater \cite{wilson2009phycodnaviridae}, and soil \cite{schulz2018hidden}. In marine ecosystems, they play key roles in the microbial food web, infecting diverse eukaryotic microorganisms and influencing host metabolism \cite{buscaglia2024adaptation, meng2021quantitative, schulz2020giant}. Notably, NCLDVs associated with algal blooms contribute to carbon export to deeper ocean layers, making them essential players in global carbon cycles and relevant to studies of carbon neutrality \cite{kaneko2021eukaryotic, kavagutti2023high, ha2023assessing}. Beyond their ecological roles, NCLDVs encode genes that confer resistance to antibiotics, potentially acting as vehicles for the transmission of antibiotic resistance genes \cite{yi2024giant}. Research shows that NCLDVs carry a higher number of ARGs on their genomes compared to other viruses \cite{chatterjee2019giant, li2024viral}, further underscoring their importance in both ecology and human health.

\label{sec:intro}

Despite their ecological and health-related significance, our understanding of NCLDVs remains limited, largely due to challenges in discovering new members of this expansive group. Historically, the identification of NCLDVs has relied on co-cultivation or isolation alongside their native hosts \cite{fischer2016giant}. However, this approach can only uncover a small fraction of their diversity, as many NCLDV hosts are difficult or impossible to cultivate. More recently, metagenomic sequencing has emerged as a powerful method for discovering new viruses, enabling the direct retrieval of genetic sequences from environmental samples without the need for cultivation. Notably, the taxonomic distribution of NCLDV Metagenome-Assembled Genomes (MAGs) differs significantly from that of genomes obtained through cultivation \cite{schulz2020giant}. As a result, metagenomics is now regarded as the primary approach for uncovering novel NCLDVs, driving rapid growth in their identification. For example, recent research \cite{buscaglia2024adaptation} recovered 1,384 NCLDV MAGs from metagenomes sampled across temperate to frigid zones, significantly broadening our understanding of the adaptation and metabolic strategies of NCLDVs. Regardless of the rapid growth, identifying NCLDV sequences remains challenging. First, NCLDV genomes are highly diverse, with Amino Acid Identity (AAI) between different families as low as 20\% \cite{moniruzzaman2020dynamic}. This extreme divergence means that novel NCLDVs in metagenomic data may share little similarity with genomes in reference databases, reducing sensitivity in computational analyses. Capturing general patterns at the amino acid level or using homology searches to identify NCLDV signatures is therefore difficult. Second, NCLDV genomes often share homologous genes with other DNA viruses, particularly bacteriophages \cite{aylward2021viralrecall}. Evidence suggests that the common ancestor of NCLDVs likely evolved from bacteriophages \cite{koonin2010origin}. Thus, the blurred boundaries between NCLDVs and bacteriophages tend to confuse the learning models. As a result, efficient identification methods are urgently needed to capture the general patterns of NCLDV genomes and establish robust decision boundaries using carefully constructed negative data.

\subsection{Related work}
\label{sec:relate}

\begin{table}[h!]
\centering
\resizebox{1\linewidth}{!}{\begin{tabular}{p{4cm}p{4cm}p{8cm}c}
\hline
Group                  & Work                                                                    & Description                              \\ \hline
\rowcolor[HTML]{EFEFEF} 
Alignment-based        & ViralRecall \cite{aylward2021viralrecall}                                                             & pHMM-based alignment against Giant Virus Orthologous Groups (GVOGs) and Pfam database            \\

\rowcolor[HTML]{EFEFEF} 
                       & Buscaglia et al. \cite{buscaglia2024adaptation}                                  & Reads mapping against known NCLDV genomes    \\

\rowcolor[HTML]{EFEFEF} 
                       & Ha et al. \cite{ha2023assessing}                   & pHMM-based alignment against the marker gene MCP          \\
\rowcolor[HTML]{EFEFEF} 
                       & Ga{\"\i}a et al. \cite{gaia2023mirusviruses}                        & pHMM-based alignment against eight hallmark genes and 149 orthologous groups                          \\
\rowcolor[HTML]{EFEFEF} 
                       & Moniruzzaman et al. \cite{moniruzzaman2020dynamic}                           & pHMM-based alignment against five marker genes: MCP, SFII, VLTF3, PolB and A32                           \\
\rowcolor[HTML]{EFEFEF} 
                       & B{\"a}ckstr{\"o}m et al. \cite{backstrom2019virus}                  &pHMM-based alignment against the DNA polymerase                           \\ \hline
Machine-learning-based & VirSorter2 \cite{guo2021virsorter2}                                                              & Features such
as protein alignment, gene density, ribosomal binding sites (RBS) are extracted to train a random forest classifier                           \\
                       & Schulz et al. \cite{schulz2020giant} & Features of gene density, 11 motifs, average spacer length and coding density are extracted to train a random forest classifier                         \\ \hline
\rowcolor[HTML]{EFEFEF} 
Combined               &Yi et al. \cite{yi2024giant}              & Use the classifier published by \cite{schulz2020giant}, detect orthologous groups and PolB gene                          \\ \hline
\end{tabular}}
\caption{Summary of works that allow NCLDV detection.}
\label{tab:tools}
\end{table}

Several attempts have been made to detect NCLDVs in metagenomic data, which can be broadly categorized into two groups: alignment-based methods and machine learning-based methods. Table \ref{tab:tools} provides a summary of these tools along with brief descriptions. Most existing approaches rely on alignment-based strategies, using manually curated marker genes as the reference databases. Commonly used marker genes include major capsid protein (MCP), B-family DNA Polymerase (PolB), A32-atpase (A32), superfamily II helicase (SFII), viral late transcription factor 3 (VLTF3), the mRNA capping enzyme (mRNAc), RNA polymerase subunits (RNAPL and RNAPS), D5 primase/helicase (D5), and ribonucleotide reductase (RNR) \cite{aylward2021viralrecall}. Among these, MCP is the most widely used, while single-copy marker genes such as A32, PolB, VLTF3, and SFII are frequently employed for detection and phylogenetic reconstruction \cite{aylward2021phylogenomic}.  However, alignment-based methods face notable limitations. Because assembled contigs from metagenomic data are often incomplete, it is unlikely that they contain the chosen marker genes. Furthermore, marker genes are not exclusive to NCLDVs; proteins from other viruses may also contain these genes, leading to ambiguous classification. To address these issues, some alignment-based methods utilize orthologous genes from NCLDVs to build more comprehensive reference databases. For instance, ViralRecall \cite{aylward2021viralrecall} constructs a database composed of RefSeq genomes and high-quality NCLDV Metagenome-Assembled Genomes (MAGs). Then genes are predicted and translated into proteins from these genomes. Clustering algorithm is applied on the protein sequences for orthologous group construction, with each group encoded by a Hidden Markov Model (HMM). HMMER \cite{eddy2011accelerated} is then employed to align query proteins against the HMM database, allowing remote homology search. While this approach improves sensitivity of detecting NCLDV contigs, its performance still heavily depends on the quality and selection of the reference database. Additionally, its computational cost increases significantly with the increase of the database size.

Machine learning-based methods aim to automatically uncover hidden patterns in genomic sequences \cite{schulz2020giant,guo2021virsorter2} and have been adapted to the identification of NCLDVs. For instance, VirSorter2 \cite{guo2021virsorter2} leverages features such as protein alignments, gene density, and ribosomal binding sites (RBS). It then trained several classifiers using data from different groups of viruses, including NCLDVs. Notably, the distribution of RBS percentages in NCLDVs differs significantly from those of other organisms, as demonstrated in their experiments \cite{guo2021virsorter2}. These features are vectorized and input into a random forest classifier for NCLDV detection. Despite its innovations, VirSorter2 has several limitations. It requires significant time and computational resources due to the large number of HMM models used in its pipeline. Additionally, VirSorter2's primary focus is on distinguishing viruses from prokaryotic and eukaryotic genomes, rather than differentiating NCLDVs from other DNA viruses. As demonstrated in our experiments, VirSorter2’s precision decreases significantly when distinguishing NCLDVs from bacteriophages. In summary, learning-based tools face several common challenges. First, we need a context-aware feature representation that can capture the gene organization among NCLDV. Second, given the huge imbalance of NCLDV and non-NCLDV data, a systematic and automatic method is needed for constructing effective negative training samples from non-NCLDV microbes. 

\subsection{Overview}
\begin{figure*}[htbp]
    \centering
    \includegraphics[width=0.9\linewidth]{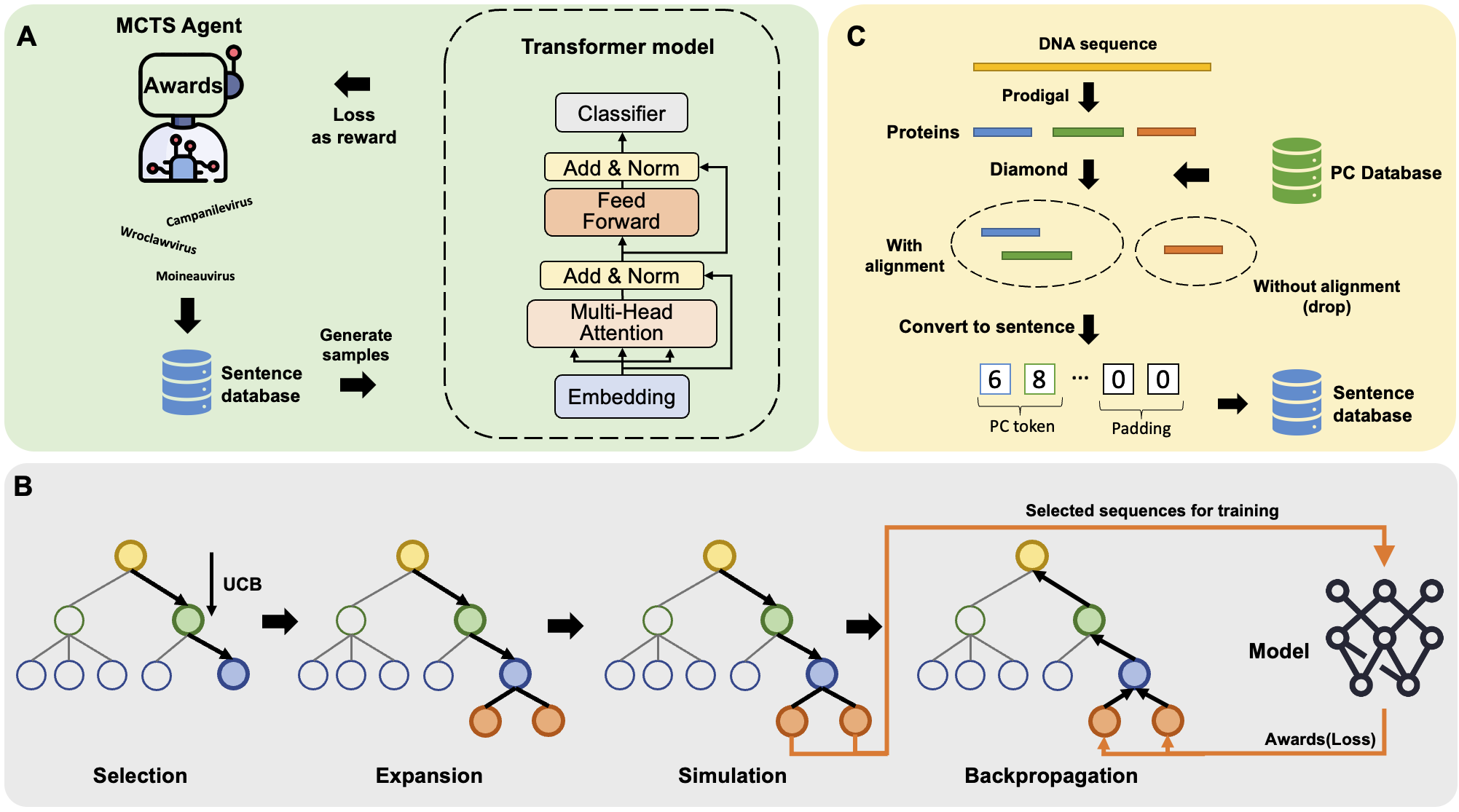}
    \caption {The Framework of GiantHunter. (A) Reinforcement Learning Process: In each iteration, the agent selects several taxa and retrieves corresponding sequences from the database. These sequences are then used to train the model. The loss values output from model serve as rewards for the agent for next selection. (B) Feature Extraction and Protein-Cluster (PC) ID Sequences: Features are extracted from nucleotide sequences via alignment to obtain Protein-Cluster (PC) ID sequences. All PC ID sequences are stored within a dedicated database. (C) Actions of Monte Carlo tree search-based agent in one iteration: each node in the taxonomy tree represents a taxon. The agent uses the Upper Confidence Bound (UCB) algorithm to iteratively select child nodes until a leaf node is reached. Child nodes are expanded, and a simulation step is performed to retrieve contigs from the database for training the model. The resulting loss values are then used to update the Monte Carlo tree via backpropagation.
}
    \label{fig:pipeline}  
\end{figure*}

In this work, we introduce a new tool GiantHunter (Fig. \ref{fig:pipeline}) for identifying NCLDV from metagenomic data. GiantHunter takes contigs or bins as input and outputs probability scores indicating the likelihood of a contig being part of NCLDV. It addresses the challenges of NCLDV detection through two innovative strategies. First, we formulate NCLDV classification as a reinforcement learning task and employ Monte Carlo tree search (MCTS) to select hard cases from a large-scale negative search space for training. NCLDV and other microbes are organized in their taxonomic tree,  where each node/taxon can represent a diverse set of genomes. MCTS is particularly well-suited for navigating such hierarchical structures because it systematically explores the tree while balancing the depth and breadth of search, ensuring efficient and comprehensive data selection. Our reinforcement learning framework allows the model to leverage the effective selection from MCTS and thus leads to a more robust decision boundary in the learning model. 
Second, GiantHunter leverages a protein-cluster-based Transformer model, where each protein cluster is used as a token in the ``language'' of NCLDV genomes. 
This adapted language model can learn protein organization and functional importance tailored specifically for NCLDVs and enhance the detection performance. 
In the experiments, we tested GiantHunter on various datasets with increasing difficulty. A comparison with two state-of-the-art methods demonstrates that GiantHunter delivers superior performance on NCLDV identification. In particular, GiantHunter improves the F1-score by approximately 10\% and reduces computational cost by 90\% compared to the second best tool. To further illustrate the utility of novel NCLDV detection, we conducted case studies on real metagenomic data from the Yangtze River. This application showcases GiantHunter’s potential for studying NCLDVs in metagenomic datasets and advancing our understanding of their ecological roles.

\section{Methods and materials}
\label{sec:method}

We formulate NCLDV identification as a reinforcement learning task to address the challenge of selecting the most informative negative training data from a large dataset. Reinforcement learning (RL) is a machine learning paradigm where an agent learns to make decisions by interacting with data and maximizing a cumulative reward from a carefully designed Transformer model. Unlike other learning strategy, RL focuses on sequential decision-making and is highly effective in scenarios where finding optimal solutions involves exploration of a large search space. RL has been increasingly applied in bioinformatics, such as protein structure prediction \cite{yang2023applying}, drug discovery \cite{popova2018deep}, and sequence optimization \cite{angermueller2019model}, where the goal is to refine models or strategies iteratively to improve performance. In our context, the database of known NCLDVs is limited, while the negative data space encompasses a vast range of other microbes. 
We thus apply RL to navigate different subsets of  non-NCLDV microbes and dynamically decide which set can lead to optimal or near-optimal NCLDV identification using the rewarding mechanisms of RL. We expect that the finally chosen negative training cases are closest to the optimal decision boundary, and thus enhancing the distinction of NCLDV from other microbes. Fig. \ref{fig:pipeline} A sketches the RL framework.

The sampling of the large number of non-NCLDV microbes in the framework of RL is conducted using Monte Carlo Tree Search (MCTS), which is widely used in RL applications for efficient decision making in large and complex search spaces. MCTS, with its tree-based sampling mechanism theoretically optimal for operating on taxonomic tree structure. In our problem, non-NCLDV microbes are organized in a taxonomic tree. MCTS is applied to navigate this tree and select taxon nodes for training. In each training iteration (Fig. \ref{fig:pipeline} A), the MCTS module will select a subset from the whole negative search space, which is then used to train a classification model. The loss value from the model serves as the reward for the MCTS agent, which directs the agent to search in the direction of higher loss values (meaning harder cases) in the search space. By iteratively selecting nodes with the greatest loss values (Fig. \ref{fig:pipeline} B), MCTS enables the model to focus on representative samples, thereby improving decision boundaries and significantly reducing false positives.

To effectively extract features from nucleotide sequences, we utilize the Transformer, a state-of-the-art language model, to automatically learn hidden patterns from the "language" of NCLDV genomes. In this implementation, sequences are treated as sentences defined by a vocabulary specific to NCLDVs (Fig. \ref{fig:pipeline} C). This approach offers two key advantages. First, certain proteins, such as the widely recognized marker gene—the major capsid protein (MCP)—play critical roles in the NCLDV life cycle or serve as strong indicators of their presence. Second, proteins often interact with one another to perform biological functions \cite{schulz2020giant}. Similar to how combinations of words form phrases with distinct meanings, specific combinations of proteins within contigs can provide important biological insights \cite{shang2022accurate}.  These considerations motivated our approach to convert contigs into protein-based sentences for downstream analysis.

In the following sections, we will first introduce the workflow of RL framework used in GiantHunter and the Monte Carlo tree search (MCTS) strategy for negative sampling. Then, we will briefly explain how we encode input sequences into PC-based sentence and detail the core parts of the Transformer model.

\subsection{Reinforcement learning framework in GiantHunter}
There are three main elements in reinforcement learning: \textit{agent}, \textit{environment}, and \textit{reward function} \cite{arulkumaran2017deep}. The environment refer to all the data we can use for training; the agent is our proposed MCTS-based search model to select nodes based on reward values; and the reward is the loss function of our Transformer model.

The environment consists of 227 complete NCLDV genomes (as positives) and 5,145 complete Caudoviricetes genomes (as negatives) downloaded from the NCBI RefSeq database.  Caudoviricetes was chosen as the negative search space in this study for two reasons. First, Caudoviricetes usually exist in NCLDV MAGs as contamination \cite{aylward2021viralrecall} and it has been speculated that the common ancestor of NCLDVs likely evolved from a bacteriophage \cite{koonin2010origin}, highlighting their genomic similarities. Second, the number of all non-NCLDV can be hundreds of thousands times more than the known NCLDVs, choosing bacteriophage as the negative search space with the domain knowledge will save a large amount of computing resources. Our experiments also reveal that the exiting methods fail to perform well on distinguishing between NCLDVs and bacteriophages, but all these methods have the ability to identify NCLDVs from prokaryotes and eukaryotes with only using bacteriophages as their negative search space. It's worth noting that, GiantHunter works better if more negative samples are included for MCTS selection as shown in the experiments.

In order to improve the robustness on short contigs, data augmentation is applied by randomly generating short DNA fragments, ranging from 5kbp to 20kbp, from the complete genomes.

During the search process, the MCTS-based agent will select a subset of negative data and feed them into the Transformer model together with positive data to get the loss value and update the weights of the Transformer model via gradient descent. Next, the agent will update the states of nodes in the Monte Carlo tree based on the reward value (the loss from the model). This process will continue for a fixed number of iterations (60,000 by default) or stop early if the performance converges.

\subsection{Negative sampling via Monte Carlo tree search}
In the MCTS-based agent, each node in the tree has the following states: the rank/level of taxonomy, the name of the taxon, the number of contigs included, the frequency of this node being visited (aka exploration), and a $Q$ value indicating the reward. Three taxonomy levels are included in our taxonomic tree: family, genus, and species. A visual root node is added to connect all family-level nodes, and the leaf nodes under specie-level nodes represent generated contigs. The $Q$ value of each node is initialized as 0.

The MCTS process includes four parts: selection, expansion, simulation, and backpropagation. The overall interaction between the MCTS agent and the Transformer-based model is shown in Fig. \ref{fig:pipeline} A. 

\paragraph{Selection} In the selection process, we begin at the virtual root and iteratively select the optimal child nodes at each level based on the node selection preference score $U_i$ as defined in Eqn. \ref{eq:ucb}. $U_i$ is a tradeoff between the node reward value (hard training cases) and node visit times (exploration). It is computed using the Upper Confidence Bound (UCB) strategy \cite{auer2000using}. 
In the beginning, the UCB strategy tends to select those unvisited nodes and then shifts to nodes with higher rewards. 

\begin{equation}\label{eq:ucb}
    U_i = Q_i + c\sqrt{\frac{\text{ln} N_i}{n_i}}
\end{equation}

\noindent $Q_i$ indicates the $Q$ value of a node during its $i$th visit. It is initialized as 0 and will be updated during the backpropagation stage. Essentially, it represents the impact level of a node on the model learning loss. Harder training cases tend to have increased $Q$ values, which will be inferred by the loss value of the Transformer-based classification model.  However, some nodes containing hard cases may consistently exhibit a large $Q$ value, restricting the agent's search to a narrow range. To mitigate this, a second term is introduced, which becomes larger for nodes with fewer visit counts, thereby encouraging the exploration of less-visited nodes.

In Eqn. \ref{eq:ucb}, $c$ represents the exploration parameter, which is a hyperparameter and set to 2 by default. A higher $c$ makes the selection favor nodes with fewer visits. $n_i$ and $N_i$ are the visit count of the current node and its parent node, respectively.  During the selection process, $U$ is calculated for all child nodes, and the node with the largest $U$ is iteratively selected until the model reaches a leaf node, which is a set of contigs corresponding to a species.

\paragraph{Expansion and simulation} Once a leaf node is selected, an expansion operation is performed by retrieving the child nodes in the taxonomy tree. If the selected node reaches the species level, no further action is needed at this step. 
As each species node has multiple leaf nodes representing all previously generated contigs, a simulation step is applied to select the contigs as the model training data.  
For each node, a fixed number of contigs are randomly sampled and passed into the Transformer model to compute their probability scores of being from NCLDV. These scores are then used to evaluate the difficulty of classification based on the true labels. Since all the contigs in this step are negative samples, their true labels are set to 0. The final cross-entropy loss is then calculated using Eqn. \ref{eq:loss}.

\begin{equation}\label{eq:loss}
    CE = \frac{1}{M}\sum_{i=1}^{M}-log(1-p_i)    
\end{equation}

\noindent where $M$ is the number of sampled contigs, and $p_i$ is the probability score of contig $i$ being from NCLDV, computed by the Transformer model. Once the cross-entropy loss is calculated for each node, the $Q$ value of the selected node by the selection procedure is set as the mean loss across the sampled contigs. The weights of the Transformer model are then updated using the gradient descent algorithm. In other words, the simulation step also serves as a training phase for the Transformer model.

\paragraph{Backpropagation} With the $Q$ values already initialized, we now describe how to update the $Q$ values of the nodes based on newly obtained loss values. Backpropagation is used to update the $Q$ values of parent nodes iteratively, all the way up to the root node.
For the leaf node, its new $Q$ value is updated on a momentum basis, whose formula is listed in Eqn. \ref{eq:q1}.

\begin{equation}\label{eq:q1}
    Q_i^l = \alpha Q' + (1-\alpha)Q_{i-1}^l
\end{equation}

\noindent $Q_i^l$ represents the $Q$ value of the leaf node during its $i$th visit, and $l$ refers to the leaf. $Q'$ is the newly calculated $Q$ value, and $\alpha$ is a tunable momentum coefficient ranging from 0 to 1, with a default value of 0.3. Because the computation of the $Q$ value involves random sampling, it can result in oscillations. To address this, the momentum-based updating method is used to smooth the $Q$ value updates. For non-leaf nodes in the tree, their $Q$ value is updated based on the weighted sum of their child nodes' $Q$ values, as shown in Eqn. \ref{eq:q2}. 

\begin{equation}\label{eq:q2}
    Q_i^p = \frac{1}{m^p}\sum_{\text{All children nodes}}m^cQ_i^c
\end{equation}

\noindent $Q_i^p$ represents the $Q$ value of the parent node during its $i$-th visit, and $p$ indicates the parent. Similarly, $Q_i^c$ refers to the $Q$ value of the child node, while $m^p$ and $m^c$ denote the contig counts of the parent and child nodes, respectively. Weighted sum allows for child nodes with more contigs to contribute more to their parent node's $Q$ value, which is more realistic.

\subsection{Transformer model}

The PC-based Transformer model has been demonstrated to be both efficient and accurate in numerous studies \cite{shang2022accurate, tang2023plasme}. Its structure, shown in Fig. \ref{fig:pipeline} A, incorporates a multi-head attention mechanism and residual connections, enabling it to effectively capture the importance and associations of PC tokens. In the following paragraphs, we will provide details on feature extraction and briefly introduce the multi-head attention mechanism.

\paragraph{Feature extraction and PC token construction}
Protein Clusters (PCs) are groups of proteins that share significant sequence similarities and are widely used to extract functional features from contigs. In this study, Prodigal \cite{hyatt2010prodigal} is employed to predict genes in contigs and translate them into proteins. Specifically, all genome sequences in the RefSeq training dataset are processed with Prodigal using default parameters to generate protein sequences. Then, an all-against-all sequence alignment of the resulting protein sequences is performed using DIAMOND BLASTP \cite{buchfink2021sensitive}. We then cluster the protein sequences into clusters using a graph-based clustering algorithm. To construct a protein-similarity graph, the software mcxload \cite{van2008graph} is used to process the BLAST results, converting edge weights from bit scores to negative log values and capping them at a maximum of 300. The Markov Clustering Algorithm (MCL) \cite{van2008graph} is then applied to this graph to identify protein clusters. All parameters for these software are provided in Supplementary Table S1. Clusters containing fewer than two proteins are discarded, resulting in a final set of 6,526 PCs.

Once we construct the PC set, we convert each input sequence into a vector of PC ID, as shown in Fig. \ref{fig:pipeline} C, which serves as an input to the Transformer model. First, the sequence is translated into protein sequences using Prodigal with the \textit{-p meta} parameter. These protein sequences are then aligned to the pre-built PC database. For each protein, the best hit is selected to assign its corresponding PC ID. Sequences that do not produce any alignment results are discarded and treated as non-NCLDV in the prediction.

\paragraph{Protein cluster-based Transformer}
The Transformer model, originally developed for Natural Language Processing (NLP), has gained popularity in biological sequence analysis due to its ability to capture contextual relationships within sequences. In this study, we use the Transformer model to process sequences of PC IDs, as certain genes in NCLDV genomes exhibit important interactions that serve as key signatures of this phylum. The maximum sentence length is set to 1,000 since 96\% of complete NCLDV genomes in RefSeq have fewer than 1,000 proteins. The model outputs the probability of whether the contigs belong to NCLDVs.

Our model has two main components: the embedding layer and the self-attention layer. The embedding layer converts discrete PC IDs into continuous embedding vectors. It includes word embedding, which maps IDs to vectors via a lookup table, and positional embedding, which encodes the order of tokens to preserve sequential information. The word and positional embeddings are summed to create the final embedding vectors, with a dimension of 768. Self-attention is the core of the Transformer model, relating different sequence positions to compute token representations \cite{vaswani2017attention}. It generates Query (Q), Key (K), and Value (V) matrices from token embeddings and uses them to identify co-occurring PCs. Multi-head attention, an extension of basic attention, captures diverse token relationships. Finally, the output of the Transformer is passed through two fully connected layers with a sigmoid activation function to predict the probability of a contig being from NCLDVs. For a more detailed explanation, please refer to the supplementary materials.

\subsection{Metrics}

As NCLDV detection is a binary classification task in machine learning, the widely used metrics for evaluating classification performance are precision, recall, and F1-score. Their formulas are listed below: 

\begin{equation}
    \label{m1}
    \text{precision} = \frac{ TP }{TP+FP}  
    \vspace*{-0.1cm}
\end{equation}

\begin{equation}
    \label{m2}
    \text{recall} = \frac{ TP }{TP+FN}  
    \vspace*{-0.1cm}
\end{equation}

\begin{equation}
    \label{m3}
    \text{F1-score} = \frac{ 2*\text{precision}*\text{recall} }{\text{precision}+\text{recall}}
\end{equation}

\noindent In NCLDV detection task, true positive (\textit{TP}), false negative (\textit{FN}), and false positive (\textit{FP}) are the number of corrected detected NCLDV contigs, the number of NCLDV contigs misclassified into non-NCLDV contigs, and the number of falsely identified NCLDV contigs, respectively. The Area Under the ROC Curve (AUCROC) is also reported for comparison.

\section{Result}
\label{sec:result}
In this section, we present the experimental results on various datasets and compare the performance of GiantHunter with existing tools. Since the field of NCLDV identification remains relatively under explored, we selected two widely accepted tools for comparison: VirSorter2 (learning-based) \cite{guo2021virsorter2} and ViralRecall (alignment-based) \cite{aylward2021viralrecall}. To establish a fair comparison, modifications were made to the configurations of the available tools. VirSorter2 was re-trained using all complete genomes of our training set where the original negative cellular sequences are replaced with phage genomes. Similarly, the reference database of ViralRecall was replaced with HMM profiles derived from the NCLDV genomes of our training set. Furthermore, we present the results of the publicly released version of VirSorter2 (with NCLDV extension) and ViralRecall for reference. In all experiments, GiantHunter was evaluated using its default score cutoff of 0.5, while the other two tools were tested with their default classification score settings.

In the following sections, we rigorous evaluate the performance of all three tools on RefSeq datasets of varying difficulty levels. Then, we demonstrate the impact of the MCTS strategy by comparing the performance of models trained on MCTS-selected datasets versus randomly sampled datasets and visualizing the top ten selected PCs (shown in Supplementary Figure S1). Finally, we illustrate the utility of GiantHunter on real sequencing data. 

\subsection{Performance on the RefSeq dataset split by time}
\label{sec:easy}
\begin{figure}[!h]
    \centering
    \includegraphics{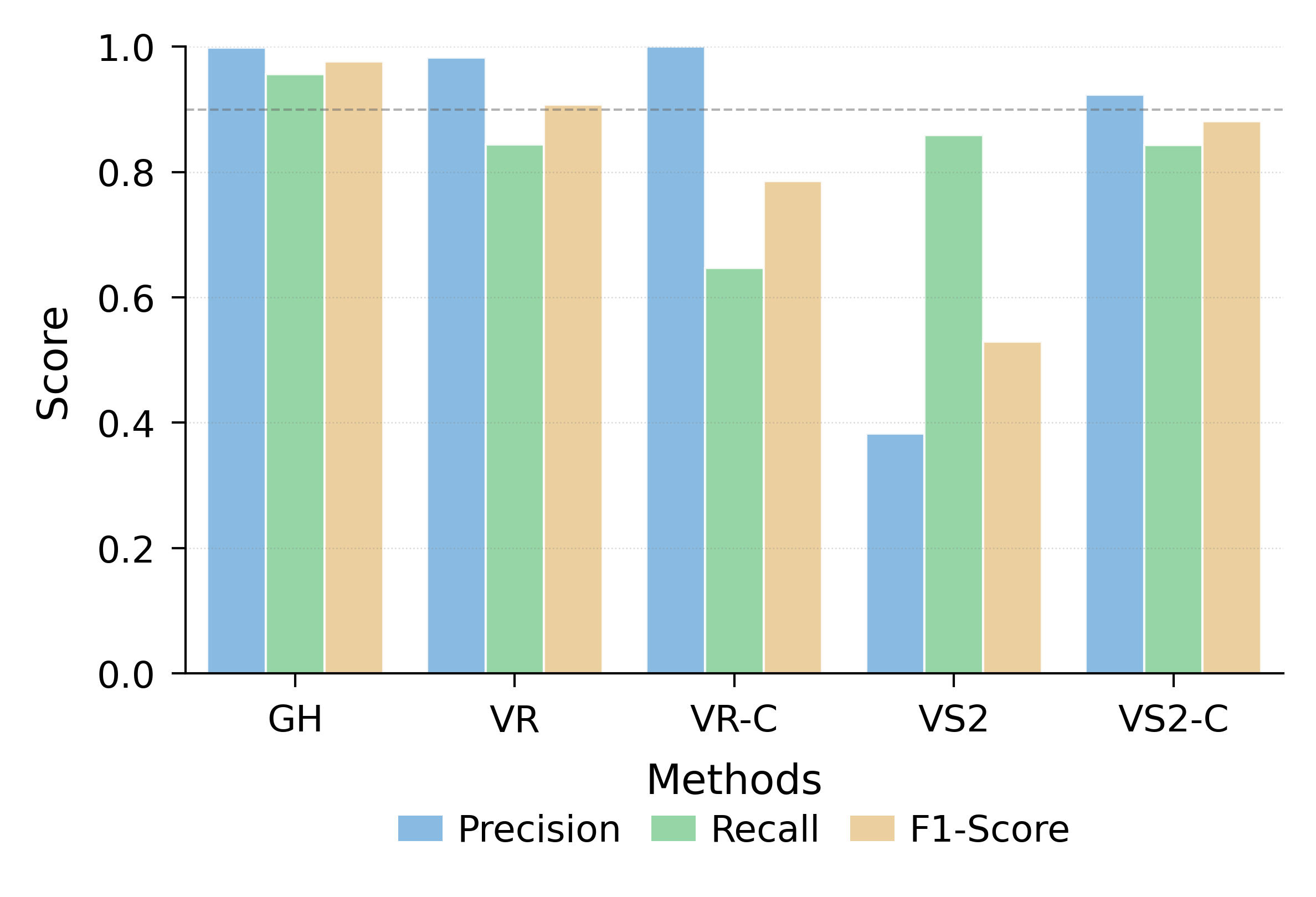}
    \caption{The performance of five models on the RefSeq dataset split by time. X-axis: Abbreviations for each tool. GS: GiantHunter; VR: ViralRecall; VR-C: ViralRecall with the customized database; VS2: VirSorter2; VS2-C: VirSorter2 with the customized database. Y-axis: Values of three metrics (precision, recall, and F1-score). }
    \label{fig:easy_all}    
\end{figure}

To create training and test sets without overlaps, we first evaluate GiantHunter on time-split data. 
Genomes released before 2018 were selected as the training set, while genomes released after 2018 were used as test sets. All genomes were fragmented into contigs of varying lengths, ranging from 5 kbp to 20 kbp, as described in the Methods section. The model was trained using the Monte Carlo Tree Search (MCTS) strategy, which focuses on selecting hard cases from the generated negative search space. After training, the model parameters were fixed and used to make predictions on the test set. The results for the complete test set are presented in Fig. \ref{fig:easy_all}.

As shown, GiantHunter consistently outperforms  other tools across all three metrics. ViralRecall achieves a high precision comparable to GiantHunter but exhibits lower recall, reflecting both the strength and limitation of alignment-based methods. Notably, replacing ViralRecall's reference database with the same training set reduces its sensitivity. Upon investigating the differences between the provided and customized databases, we found this discrepancy may be caused by potential data leakage, as the provided database appears to contain some NCLDVs from our test set. In contrast, re-training VirSorter2 with our customized training data significantly improves its performance. This is likely because the original negative data in VirSorter2, most of which are prokaryotes/eukaryotes, was less suitable for NCLDV detection task. These results highlight the importance of selecting appropriate negative data tailored to the specific problem.

\begin{figure}[htbp]%
\centering
\includegraphics[width=0.6\linewidth]{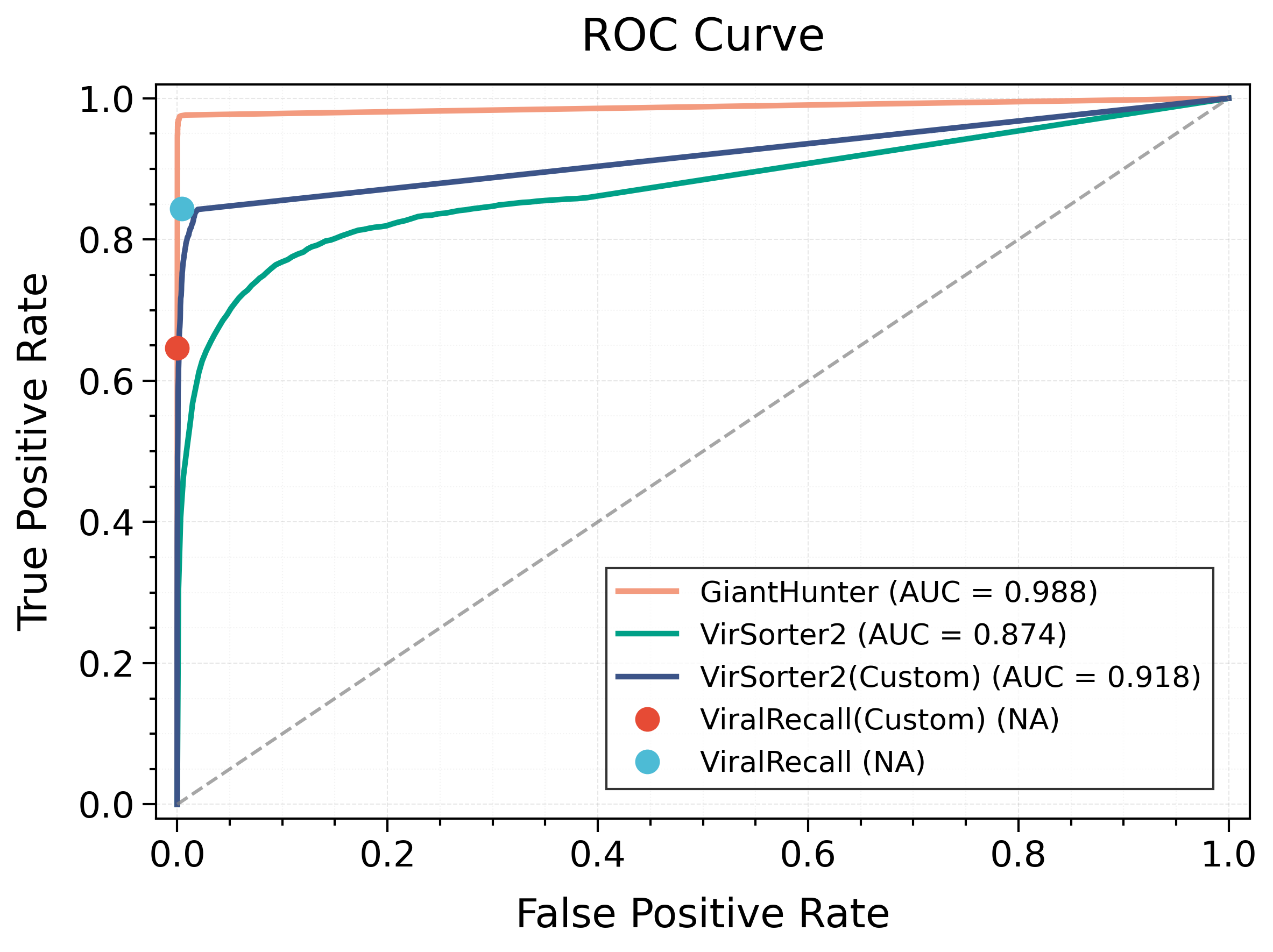}
\caption{The ROC curves of the binary virus classification on the test data by different tools. The number following the tool name is the value of the AUCROC. ViralRecall does not output a probability score for the prediction, and thus only TPR and FPR are reported.}
\label{fig:easy_roc}
\end{figure}

ROC curve was generated using the output scores of each tool, as shown in Fig. \ref{fig:easy_roc}. The area under the ROC curve (AUC) indicates that GiantHunter produces more reliable results across all contigs. Additionally, we compared the default models with the customized models trained on our training set for both VirSorter2 and ViralRecall. Since ViralRecall does not output probability scores, its performance is represented as a single point on the curve. For VirSorter2, re-training the model significantly improves its performance. In contrast, for ViralRecall, the customized version reduces its performance, highlighting the potential impact of different database constructions.

To assess the effectiveness of the Monte Carlo Tree Search (MCTS) strategy, we conducted an ablation study. To minimize the influence of randomness during sampling, we fixed the random seed. Models were then trained using either MCTS or random sampling for the same 60,000 steps. Their performance was evaluated on the validation set using precision, recall, and F1-score as metrics. The results are summarized in Fig. \ref{tab:abs}.

\begin{table}[htbp]
\centering
\begin{tabular}{p{1.5cm}p{1.9cm}p{1.9cm}p{1.9cm}}
\hline
       & \textbf{Precision} & \textbf{Recall} & \textbf{F1-score} \\ \hline
MCTS & 0.968              & 0.901           & 0.933             \\
Random   & 0.956              & 0.877           & 0.915             \\ \hline
\end{tabular}
\caption{The performance of models trained with datasets sampled by different methods: MCTS and random sampling.}
\label{tab:abs}
\end{table}

Across all three metrics, the model trained with the MCTS strategy outperforms the one trained with random sampling. By leveraging MCTS, hard negative instances in the training set are effectively identified and prioritized during training. This allows the model to establish a more robust decision boundary for the NCLDV identification task. In addition, the gaps become larger ($\geq$5\%), if more negative samples, such as prokaryotes and eukaryotes, are included for training.

\subsection{Performance on the low-similarity dataset}
Predicting contigs that are less similar to those in the training set is more challenging. To evaluate the ability of discover diverse NCLDVs from metagenomic data, we used genome distance as a similarity metric and control the maximum similarity between the training and test sets. Inspired by \cite{guan2023phagenus}, a clustering-based approach VIRIDIC \cite{moraru2020viridic} was employed to partition the dataset. Specifically, we computed distances between complete genomes and distance matrices are clustered using the single-linkage mode. This ensure that the maximum similarity between clusters does not exceed 30\%. Clusters are then sorted by size, and genomes from each cluster are sequentially added to the training set and test set.

\begin{figure}[htbp]
    \centering
    \includegraphics[width=0.5\linewidth]{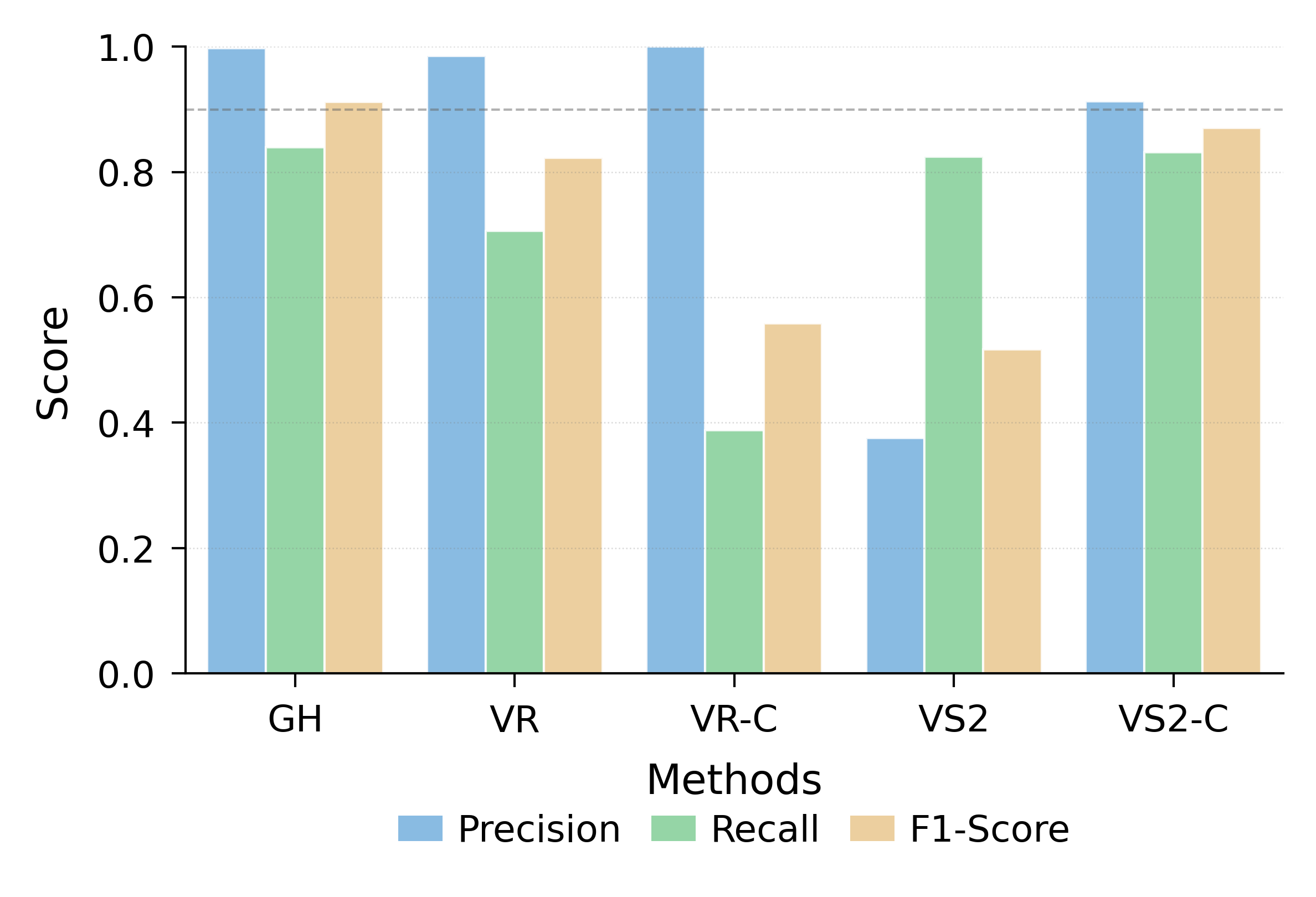}
    \caption{The performance of five models on the RefSeq dataset split by similarity. X-axis: Abbreviations for each tool. GS: GiantHunter; VR: ViralRecall; VR-C: ViralRecall with the customized database; VS2: VirSorter2; VS2-C: VirSorter2 with the customized classifier. Y-axis: Values of three metrics precision, recall and F1-score.}
    \label{fig:hard_all}    
\end{figure}

The performance of GiantHunter, VirSorter2, and ViralRecall on the low-similarity dataset is presented in Fig. \ref{fig:hard_all}. As expected, the results show trends similar to those observed in Section \ref{sec:easy}. While the recall of GiantHunter decreases by 7\%, the precision remains high, even as the dataset becomes more challenging. Despite this, GiantHunter continues to outperform all other tools, demonstrating its effectiveness in identifying diverse NCLDVs.

\subsection{Performance on sequences with different lengths}
Considering that metagenomic assembly often produces incomplete NCLDV sequences, we evaluated the performance on simulated contigs of varying lengths, as shown in Fig. \ref{fig:length}. Detailed precision, recall, and F1-scores for each tool and group are provided in Supplementary Tables S2 and S3.

\begin{figure}[H]
    \centering
    \includegraphics[width=0.6\linewidth]{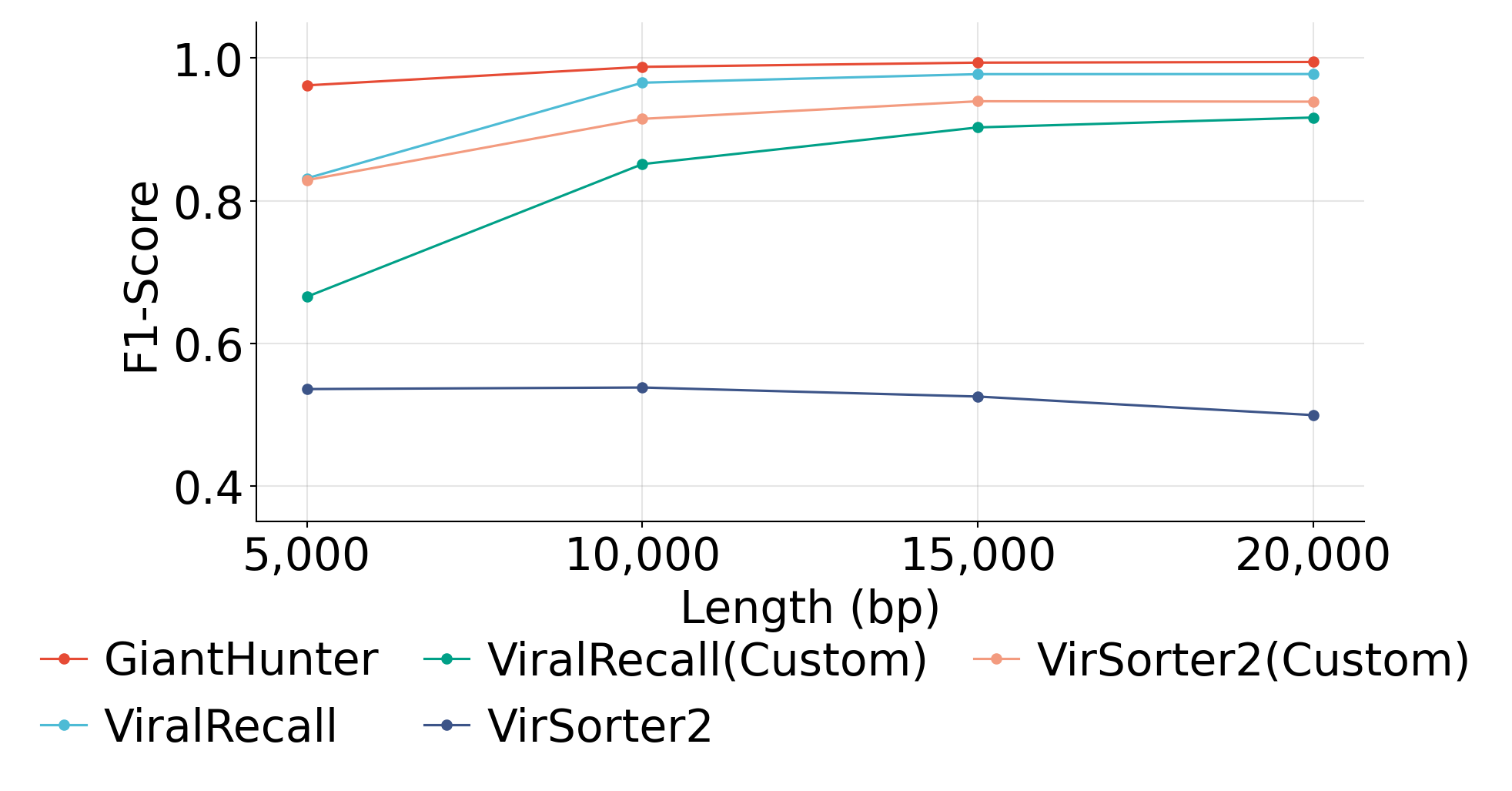}
    \caption{The per-length classification performance on the date-split dataset. X-axis: The length of contigs. Y-axis: F1-score}
    \label{fig:length}  
\end{figure}

Across all contig lengths, GiantHunter demonstrates relatively consistent and high performance compared to the other two tools. The F1-scores of most of methods increase with contig length, as longer contigs typically provide more evidence for classification, thereby improving model accuracy. However, we observed that while recall increases with contig length, the precision of VirSorter2 decreases significantly. This may be because VirSorter2 was originally designed to distinguish viruses from other microbes, and its selected features may not be well-suited for differentiating NCLDVs from other viruses. As contig length increases, these ambiguous mapped features appear to confuse the model, leading to the misclassification of some phages as NCLDVs and causing a drop in precision. Notably, for short contigs of 5 kbp, GiantHunter achieves significantly higher accuracy than the other two tools, underscoring its potential to recruit more NCLDV sequences from metagenomic data and enhance subsequent binning processes. 


\subsection{Running time comparison}
In the inference process, because the MCTS and the model are fixed, the most resource-intensive components of GiantHunter are the gene finding step and sequence alignment. These steps convert input sequences into protein cluster-based sentences, which are then processed by the Transformer model. Since metagenomic sequencing typically generates a large volume of sequences, efficiency is critical for NCLDV analysis. To benchmark performance, we measured the elapsed time required to classify 10,000 NCLDV sequences for each tool.


\begin{table}[htbp]
\centering
\begin{tabular}{cccc}
\hline
\textbf{Program}  & \textbf{GiantHunter} & \textbf{ViralRecall} & \textbf{VirSorter2} \\ \hline
Time(min) / 10,000 contigs & 8                   & 260                &  230                   \\ \hline
\end{tabular}
\label{time}
\caption{The elapsed time to make predictions for the RefSeq test genomes split by date. All the methods are run on Intel(R) Xeon(R) CPU E5-2670 v2 @ 2.50GHz CPU with eight cores.}
\end{table}

As shown in Table 3, GiantHunter is significantly faster than both ViralRecall and VirSorter2. This is primarily due to the reliance of ViralRecall and VirSorter2 on numerous HMMER models for aligning proteins to reference databases, which is a computationally expensive process. Additionally, ViralRecall employs a sliding window approach to detect subsequences within contigs, analyzing fragments and integrating results for classification. Therefore, GiantHunter can achieve optimal performance while maintaining high efficiency.



\begin{figure*}[htbp]
    \centering
    \includegraphics[width=\linewidth]{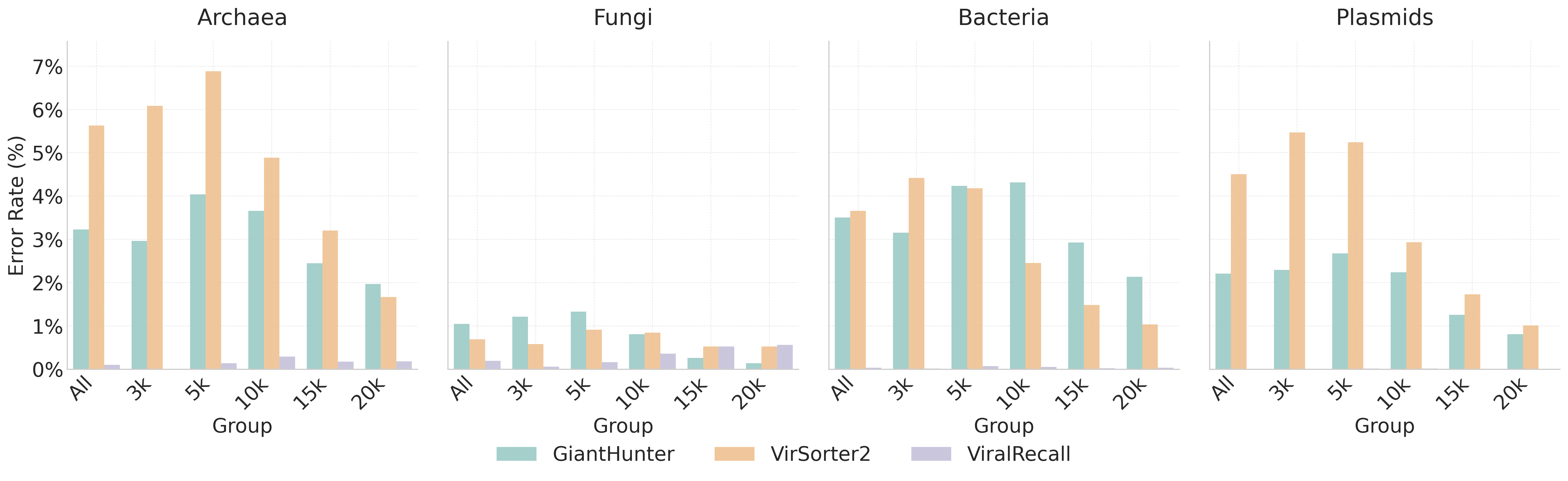}
    \caption{The error rates of three tools on four groups of data. X-axis: The length of contigs; ``All'' represents the complete genomes. Y-axis: Error rates.}
    \label{fig:ood}   
\end{figure*}

\subsection{Performance on out-of-distribution data}
Real metagenomic data contain contigs from a wide range of microorganisms, including archaea, bacteria, fungi, and plasmids. Because our negative training data only include phage contigs, it is essential to validate the performance of the models on these out-of-distribution (OOD) data. To this end, we downloaded genomes of archaea, bacteria, and fungi from the NCBI RefSeq database released after January 2024. Plasmids were obtained from \cite{brooks2019curated}. Due to the lengthy inference times required by ViralRecall and VirSorter2, we randomly sampled a subset of genomes per domain, adjusting the sampling ratios to account for their varying genome lengths. All genomes were then fragmented into contigs of lengths 3 kbp, 5 kbp, 10 kbp, 15 kbp, and 20 kbp. In total, we obtained 27,105, 184,962, 46,090, and 48,109 contigs for archaea, bacteria, fungi, and plasmids, respectively. For ViralRecall and VirSorter2, default databases, models, and parameters were used. 
Since the purpose of this experiment is to mimic realistic usage scenarios where users do not re-train the model, the customized versions of the two tools are not included. Error rates on all contigs, as well as on groups of different lengths, are presented in Fig. \ref{fig:ood}.

For all domains, ViralRecall achieves the lowest error rates. this is expected because ViralRecall has a strict alignment strategy and lead to a high precision and low sensitivity as shown in the previous experiments. Among the two learning-based tools, GiantHunter generally exhibits lower error rates than VirSorter2. One plausible reason of a higher error rates in bacteria and archaea of GiantHunter may be caused by the presence of prophages. Further investigation reveals that bacterial sequences mistakenly classified as NCLDVs frequently usually contain prophages indeed, which may confuse the model. It is worth noting that the error rates drop if the misclassified sequences in this experiment are used as negative samples and has very slightly impact on the previous NCLDV detection performance.


\subsection{Case study: Yangtze river}
After rigorously validating GiantHunter, we applied it to analyze metagenomic data from the Yangtze River, the longest river in China and the third longest globally. A total of 211 samples were collected by co-author Dr. Wang from January to October 2020, spanning river segments across three regions defined by their proximity to the Three Gorges Dam and urban reaches. Six cities located upstream and downstream of the dam were collected in this study, yielding 60 metagenomic datasets sequenced in June. Previous work demonstrated that the Three Gorges Dam significantly influences aquatic ecosystems, particularly the composition of eukaryotic microbial communities \cite{tan2022microbial, wang2023dam}. Consequently, we hypothesized that the diversity of NCLDVs may be shaped by changes in host community dynamics.

First, we grouped the samples by their location and applied cross-sample assembly using MEGAHIT \cite{li2015megahit, li2016megahit} with the default settings, Second, we fed all these contigs into GiantHunter and identified 201,153 candidate NCLDV contigs. Third, cross-sample abundance and correlation information were utilized for binning using MetaBAT v2 \cite{kang2019metabat}. To assess the quality of the NCLDV MAGs, CheckV v1.0.3 (database v1.5) \cite{nayfach2021checkv} was used to filter MAGs with an estimated completeness of $\le$50\%. This process retained 298 MAGs. We further validated these MAGs using the seven-gene markers proposed in \cite{aylward2021phylogenomic}. Out of the 298 MAGs, 286 contained at least one marker gene. The proportion of MAGs containing all seven markers was approximately 58\% in our study, compared to ~70\% reported in the NCLDV database collected by \cite{aylward2021phylogenomic}. This result indicates that relying solely on marker genes for NCLDV identification may hinder the discovery of the full diversity of these viruses.

To further analyze the diversity of NCLDVs across different cities, BBTools \cite{bushnell2014bbmap} was used to summarize alignment statistics. MAGs with coverage $\geq$85\% and depth $\geq$1x were considered to "exist" in a given sample. The alpha diversity of NCLDVs in each city was then calculated based on the RPKM values generated by BBTools.

\begin{figure}[htbp]
    \centering
    \includegraphics[width=0.6\linewidth]{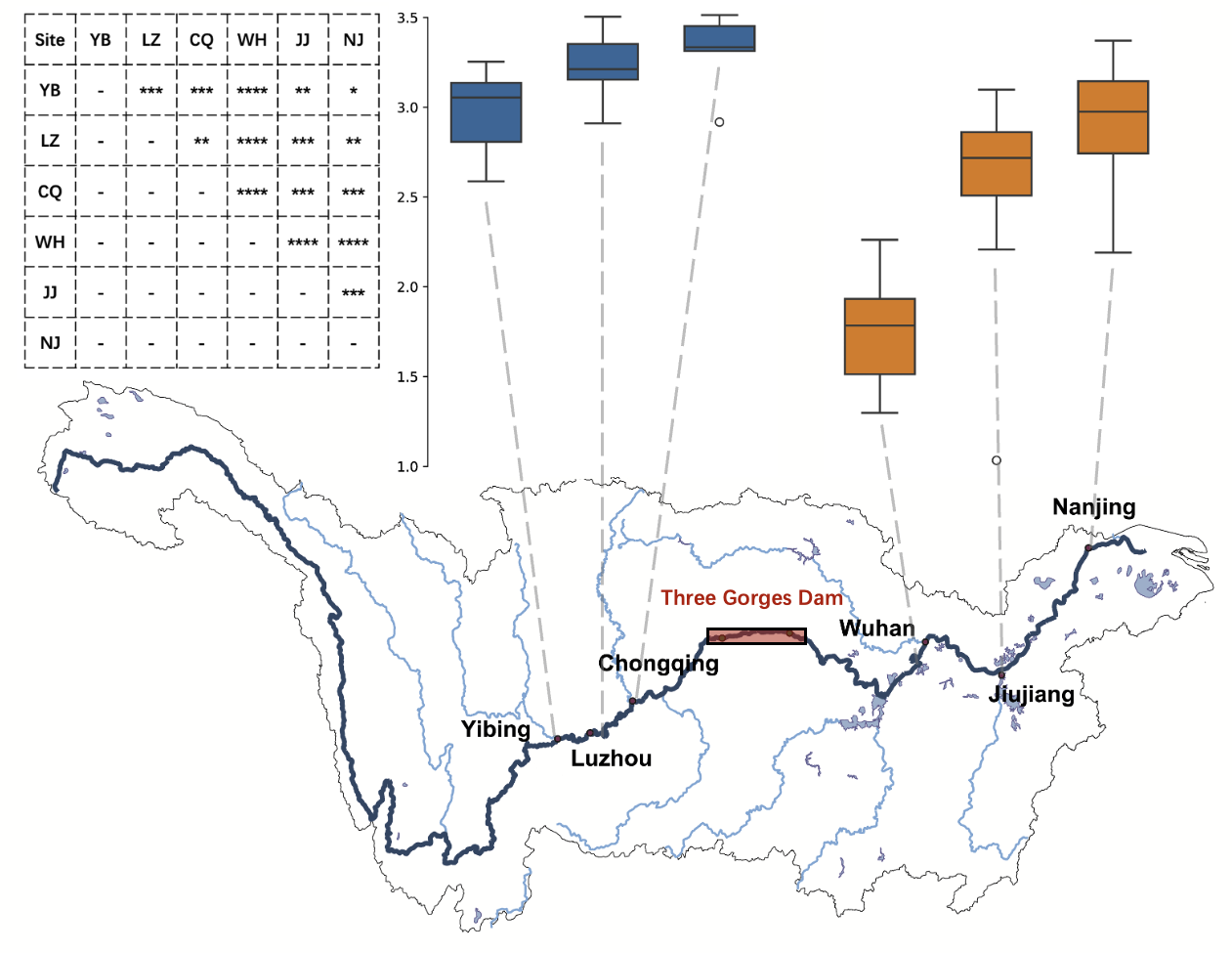}
    \caption{Alpha diversity of NCLDV in different cities along the Yangtze River and statistical significance test between cities. In the table, '*' means 1e-1, '**' means 1e-2, and etc.}
    \label{fig:Yangtze}    
\end{figure}

As shown in Fig. \ref{fig:Yangtze}, the Three Gorges Dam significantly affects the composition and diversity of NCLDV communities. Upstream, NCLDV diversity increases with proximity to the dam. A plausible explanation is that the dam slows the flow velocity, creating more stable conditions that enable the establishment of diverse microbial communities \cite{wang2023dam, wang2021lifestyle}. Notably, a significant drop in diversity is observed in the first downstream city, Wuhan. This is likely a consequence of the Three Gorges Dam altering downstream flow regimes, leading to reduced sediment transport, changes in nutrient availability, or lower turbidity. These changes may reduce the availability of hosts for NCLDVs, resulting in decreased diversity. Further downstream, NCLDV diversity begins to recover. As the river flows farther from the immediate impacts of the dam, environmental conditions stabilize, allowing diversity to return to levels comparable to those observed upstream. This study demonstrates that GiantHunter is a robust tool for NCLDV detection and highlights its potential to advance our understanding of newly identified NCLDVs and their ecological roles.

\section{Discussion}

In this work, we present GiantHunter, a reinforcement learning framework that leverages MCTS and a protein cluster-based Transformer model for NCLDV detection. The MCTS sampling method efficiently selects hard cases from negative search space for training, while the Transformer model captures the significance and associations between proteins to establish a more reliable decision boundary. Our rigorous experiments demonstrate that GiantHunter achieves robust performance, outperforming all other methods in NCLDV detection. Additionally, our case study using GiantHunter to analyze metagenomic sequencing data yielded valuable insights into NCLDV discovery.

Despite the significant improvements in accuracy and speed achieved by GiantHunter, there are several opportunities for optimization in future work. First, as highlighted in our experiments, only bacteriophages were included in the negative dataset. While the results are promising, occasional misclassified bacterial genomes indicates that incorporating representative bacterial genomes could further enhance performance. To address this, we plan to conduct a comprehensive MCTS on all potential prokaryotic and eukaryotic genomes present in metagenomic data and release an extended version of GiantHunter. Second, beyond NCLDV detection, our MCTS-based reinforcement learning framework has broader potential applications. It is well-suited for any classification problem involving large-scale hierarchical data. In the future, we aim to explore its application in other areas to further advance the field of bioinformatics.

%
%


\section*{Funding}
This work was supported by the Hong Kong Research Grants Council (RGC) General Research Fund (GRF) [11209823],  and the City University of Hong Kong.


\bibliographystyle{unsrt}  
\bibliography{references}

\end{document}